\documentclass[aps,prl,twocolumn,superscriptaddress]{revtex4-2}

\usepackage{graphicx}
\usepackage{upgreek}

\begin{document}

\title{Superballistic transport of thermal photons in confined many-body systems}

\author{Jian Dong}
\affiliation{Institute of Frontier and Interdisciplinary Science, Shandong University, Qingdao, Shandong 266237, China}

\author{Junming Zhao}\thanks{1}\email{jmzhao@hit.edu.cn}
\affiliation{School of Energy Science and Engineering, Harbin Institute of Technology, Harbin 150001, China}

\author{Philippe Ben-Abdallah}\thanks{2}\email{pba@institutoptique.fr}
\affiliation{Laboratoire Charles Fabry, UMR 8501, Institut d'Optique, CNRS, Universit\'{e} Paris-Saclay, 2 Avenue Augustin Fresnel, 91127 Palaiseau Cedex, France}

\author{Linhua Liu}
\affiliation{Institute of Frontier and Interdisciplinary Science, Shandong University, Qingdao, Shandong 266237, China}

\date{\today}

\begin{abstract}
Ballistic transport, realized when the system size is smaller than the mean free path of energy carriers, is traditionally regarded as the ultimate limit for energy transfer. Here, we predict a superballistic radiative heat transport regime that surpasses this limit in dilute chains of plasmonic nanoparticles confined within cavities. This anomalous regime exhibits superlinear scaling of the effective thermal conductivity (\(\kappa \sim L^{1.5}\)) and originates from the amplification of long-range interactions mediated by cavity-guided modes. Our results establish a framework for ultrafast photonic heat transport and open pathways for thermal management, information processing and energy transfer in quantum and nanoscale systems.
\end{abstract}

\maketitle
Ballistic transport is widely regarded as the ultimate limit of energy transfer, realized when the system size is smaller than the mean free path of the carriers. In bulk materials, heat conduction is governed by normal diffusion, where frequent scattering events render the thermal conductivity size independent. By contrast, in low-dimensional (LD) or nanoscale systems, the reduced phase space for scattering allows carriers to propagate over extended distances, giving rise to superdiffusive or ballistic transport regimes~\cite{Lepri2003,Gu2018,ChenG2021,Dhar2008}.
In such LD systems, anomalous heat transport can emerge, leading to a diverging effective thermal conductivity. In 1D systems this conductivity  scales with the system length as $\kappa \sim L^\alpha$ in 1D systems, where the exponent $\alpha$ depends on the specific system~\cite{Narayan2002,LiB2003,Mingo2005} and . The ballistic limit corresponds to $\alpha = 1$, describing scattering-free propagation. In 2D systems, the divergence is weaker, with a logarithmic scaling $\kappa \sim \log L$~\cite{WangL2012,XuX2014}. These scaling laws reflect the progressive breakdown of normal diffusion as dimensionality is reduced.
\begin{figure}
\includegraphics[width=7cm]{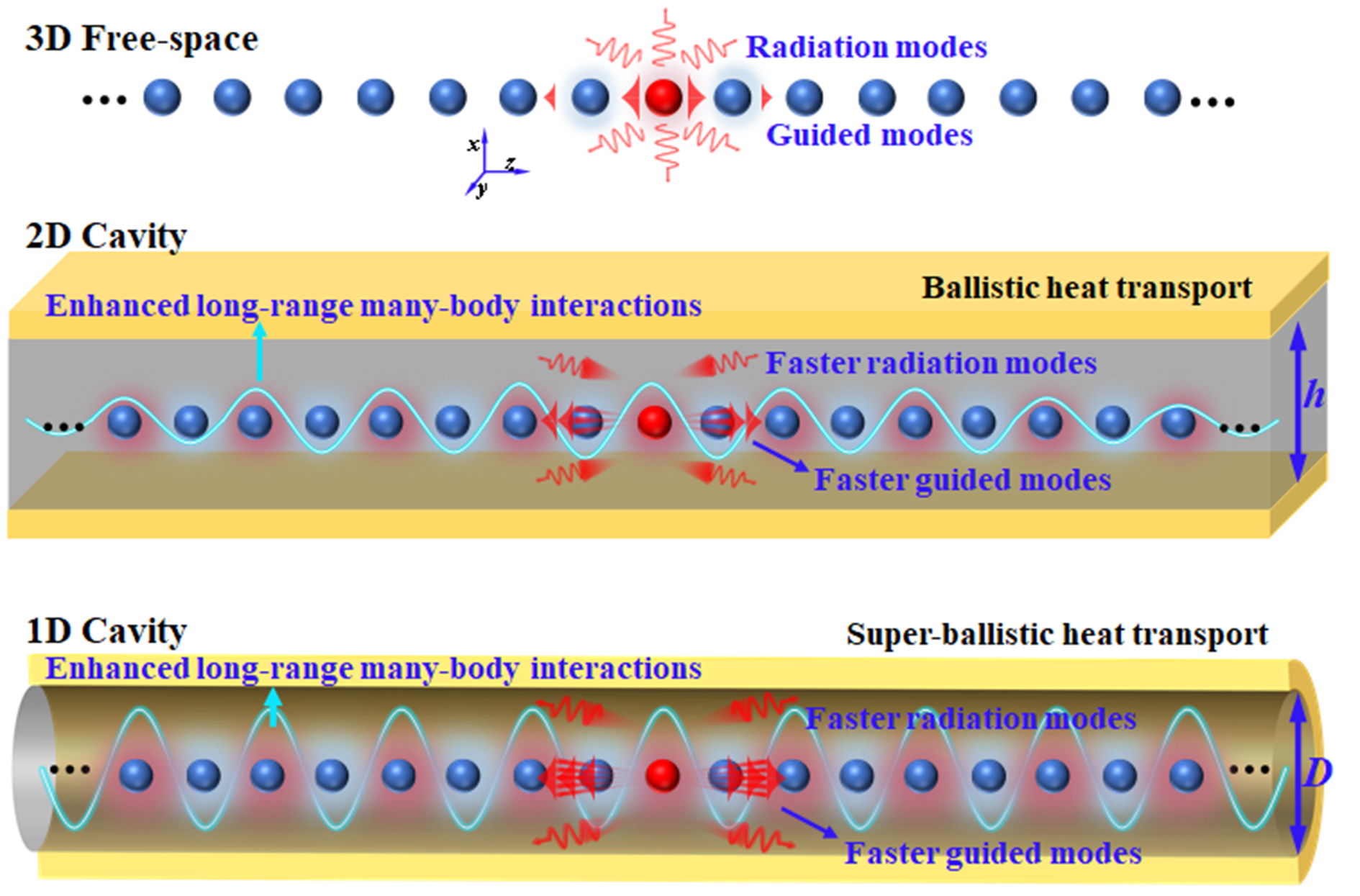}
\caption{Sketch of a plasmonic nanoparticle chain in free space, in a planar cavity or in a cylindrical cavity. Reducing the photonic dimensionality strengthens long-range many-body interactions along the chain, enabling the cooperative formation of accelerated guided and radiative modes that drive ballistic and superballistic transport.}
\end{figure}
Anomalous radiative heat transport in many-body systems has also been investigated. In particular, it has been shown that long-range near-field cooperative interactions in nanoparticle networks can induce superdiffusive heat transport governed by Lévy-like dynamics~\cite{PBA2013}. Subsequent studies demonstrated that strongly interacting many-body configurations can exhibit a crossover toward ballistic radiative transport~\cite{Latella}.

Recently, transport regimes exceeding the conventional ballistic limit—termed superballistic or hyperdiffusive—have been reported in a broad range of systems involving both bosonic and fermionic carriers. For instance, electrons flowing through viscous constrictions can collectively surpass the ballistic conductance limit due to hydrodynamic effects~\cite{Krishna2017,GuoH2017}. Superballistic spreading of photons has been observed in dynamically evolving disordered media~\cite{Levi2012}. Moreover, transient superballistic behavior has been predicted in one-dimensional disordered lattices for electrons~\cite{Hufnagel2001,ZhangZ2012} and photons~\cite{SSimon2013}, where constructive interference enhances transport beyond ballistic scaling. Similarly, phonons driven by successive heat pulses in nanoscale films have been shown to exhibit transient superballistic propagation~\cite{TangD2017}.
\begin{figure*}
\includegraphics[width=\textwidth]{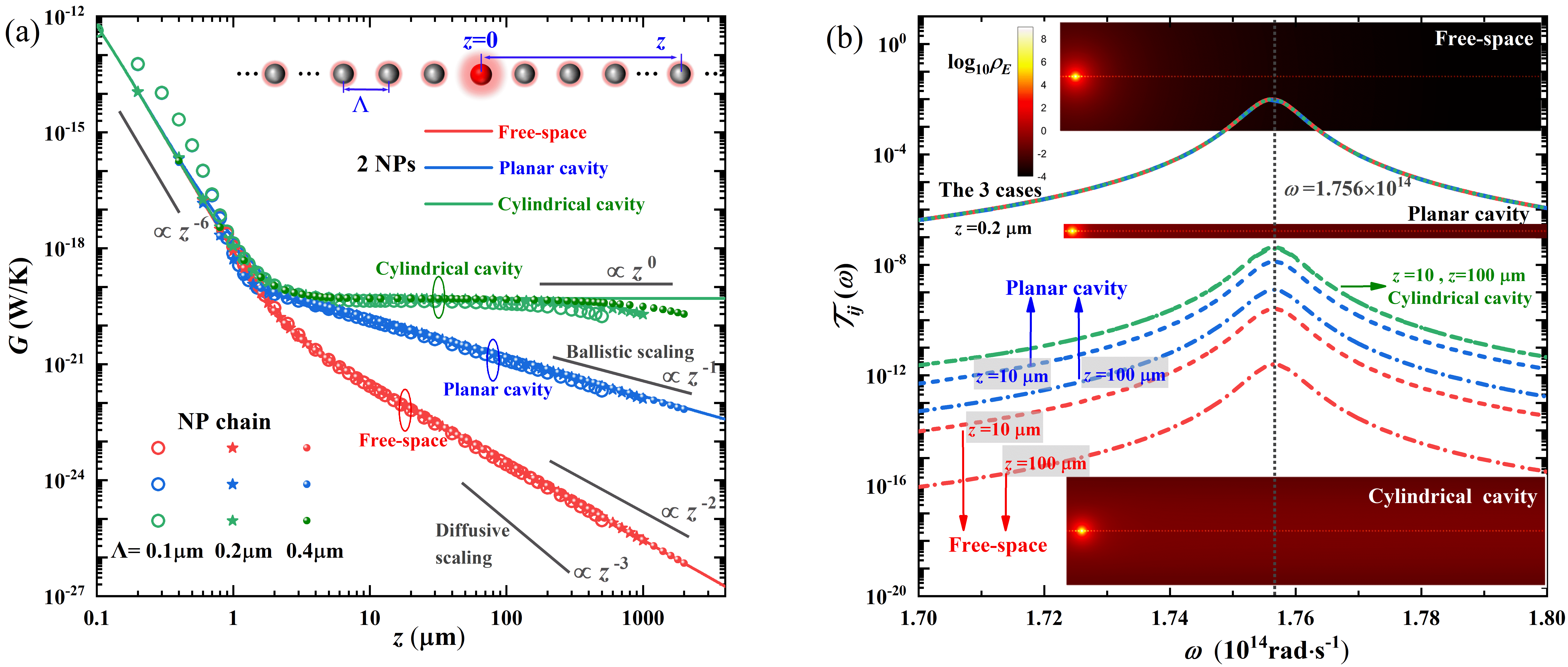}
\caption{Radiative transport along a chain of $N = 10{,}001$ SiC nanoparticles in free space or confined in planar or cylindrical cavity. (a) Spatial scaling of the thermal conductance $G$ between the central NP and other NPs at position $z$ (symbols), shown for chains with different lattice constants $\Lambda$; for comparison, $G$ between two isolated NPs is plotted as solid lines. (b) Transmission coefficient $\mathcal{T}_{ij}$ between NPs at various separations $z$ for a chain with $\Lambda = 0.2\,\upmu\mathrm{m}$. Insets display the spatial distribution of the electric component of the local density of states of electromagnetic field when only the central NP is thermally excited.}
\end{figure*}

In this Letter, we predict a superballistic heat transport regime mediated by thermal photons in confined low-dimensional many-body systems. We demonstrate that photonic confinement induces long-range interactions along the system via coupling to cavity-guided modes. These enhanced interactions promote the formation of faster, spatially delocalized transport eigenmodes, giving rise to ballistic or superballistic regimes depending on the geometric configuration. Within a generalized random-walk framework, we identify these anomalous regimes and substantiate them by demonstrating linear and superlinear scaling of the effective thermal conductivity with system length. Our findings establish a steady-state photonic counterpart to the superballistic electron flow observed through a point contact.

We consider a periodic chain of SiC nanoparticles (NPs) of radius 
$R=25\,\mathrm{nm}$ and lattice constant $\Lambda$, embedded either in free space (3D), in the midplane of a planar cavity (2D) with subwavelength gap $h=1\,\upmu\mathrm{m}$, or along the axis of a cylindrical cavity (1D) of inner diameter $D=7\,\upmu\mathrm{m}$. These particles support a surface phonon polariton which is excited at ambiant temperature~\cite{Palik}.
For a sufficiently long chain close to thermal equilibrium, local energy conservation is governed by a Chapman–Kolmogorov master equation~\cite{PBA2013,Suppl}
\begin{equation}
\frac{d\Delta T\left( {{\mathbf{r}}_{i}},t \right)}{dt}=\int_{{{R}^{d}}}^{{}}{p\left( {{\mathbf{r}}_{i}},\mathbf{r} \right)\frac{\Delta T\left( \mathbf{r},t \right)}{\tau \left( \mathbf{r} \right)}d\mathbf{r}}
\end{equation}
which describes the thermal energy $\Delta T\left( \mathbf{r},t \right)=T\left( \mathbf{r}_{i},t \right)-T_{\text{b}}$ dynamics of the system as a generalized random walk, $T\left( {{\mathbf{r}}_{i}},t \right)$ is the local temperature of the NPs, $T_{\text{b}}=300$K is the background temperature, $p\left(\mathbf{r}_{i},\mathbf{r}\right)$ is the jumps probability distribution function (PDF) defined as $p\left( \mathbf{r},\mathbf{{r}'} \right)=p\left( z-{z}' \right)={\tau \left( \mathbf{r} \right)G\left( \mathbf{r}-\mathbf{{r}'} \right)}/{C\Delta V}\;$ at a time rate $\tau {{\left( \mathbf{r} \right)}^{-1}}={\int_{{}}^{{}}{dzG\left( \mathbf{r}-\mathbf{{r}'} \right)}}/{C\Delta V}\;$ \cite{PBA2013,Suppl}. $G\left( \mathbf{r}-\mathbf{{r}'} \right)=G\left( \mathbf{r},\mathbf{{r}'} \right)$ is the radiative thermal conductance defined as~\cite{PBA2011,Biehs2021}
\begin{equation}
G\left( {{\mathbf{r}}_{i}},{{\mathbf{r}}_{j}} \right)=3\int_{0}^{\infty }{\frac{d\omega }{2\pi }{{\left. {{\mathcal{T}}_{ij}}\left( \omega  \right)\frac{\partial \Theta \left( \omega ,T \right)}{\partial T} \right|}_{T={{T}_{\text{b}}}}}}
\end{equation}
with $\mathcal{T}_{ij}$ the Landauer transmission coefficient which can be calculated from the full dyadic Green tensor~\cite{Suppl} and where 
$\Theta(\omega,T)=\hbar\omega[\exp(\frac{\hbar\omega}{k_{\mathrm B}T})-1]^{-1}$  is the mean energy of a harmonic oscillator at temperature $T$. Hence, the thermal conductance inherits the same spatial scaling as the PDF.

For random walks in one-dimensional systems governed by Lévy flights~\cite{Klages2008}, the probability density function (PDF) exhibits an algebraic decay $
p(z) \sim \frac{1}{|z|^{1+\alpha}}$ with \(0<\alpha\le 2\). Transport is diffusive in the Gaussian limit \(\alpha=2\) and approaches the ballistic regime as \(\alpha \to 0\), where increasingly long-range jumps dominate the dynamics.  
Figure~2(a) displays the spatial scaling of the radiative thermal conductance \(G\) between the central NP and the remaining particles in a chain of \(N=10\,001\) NPs. For comparison, we also show the conductance between two isolated NPs in the same configurations~\cite{Asheichyk2023,Asheichyk2018}. In the deep near-field, all three geometries exhibit a similar decay governed by short-range evanescent coupling.  
At larger separations \(z\), however, their asymptotic behaviors differ markedly. In free space, \(G\) decays as \(G \sim z^{-2}\), corresponding to a superdiffusive transport regime, as reported in~\cite{PBA2013}. When confined in a two-dimensional planar cavity, the conductance follows a slower algebraic decay \(G \sim z^{-1}\) over distances ranging from micrometers to millimeters, indicative of long-range ballistic heat transport along the chain, even in dilute configurations. Strikingly, in a one-dimensional cylindrical cavity, the radiative heat transfer between two NPs becomes nearly independent of their separation for \(z \gtrsim 1\,\upmu\mathrm{m}\), yielding an approximate scaling \(G \sim z^{0}\).  
Within the Lévy-flight framework, this behavior corresponds to an effective exponent \(\alpha \le 0\), signaling a superballistic transport regime dominated by extremely long-range interactions that surpass the conventional ballistic limit. At very large separations, small deviations from the algebraic scaling appear, which can be attributed to collective extinction effects along the chain or to a gradual crossover toward standard far-field decay. Simulations of sparser chains further confirm that reducing the interparticle density extends the distance range over which the algebraic scaling persists.
\begin{figure*}
\includegraphics[width=\textwidth]{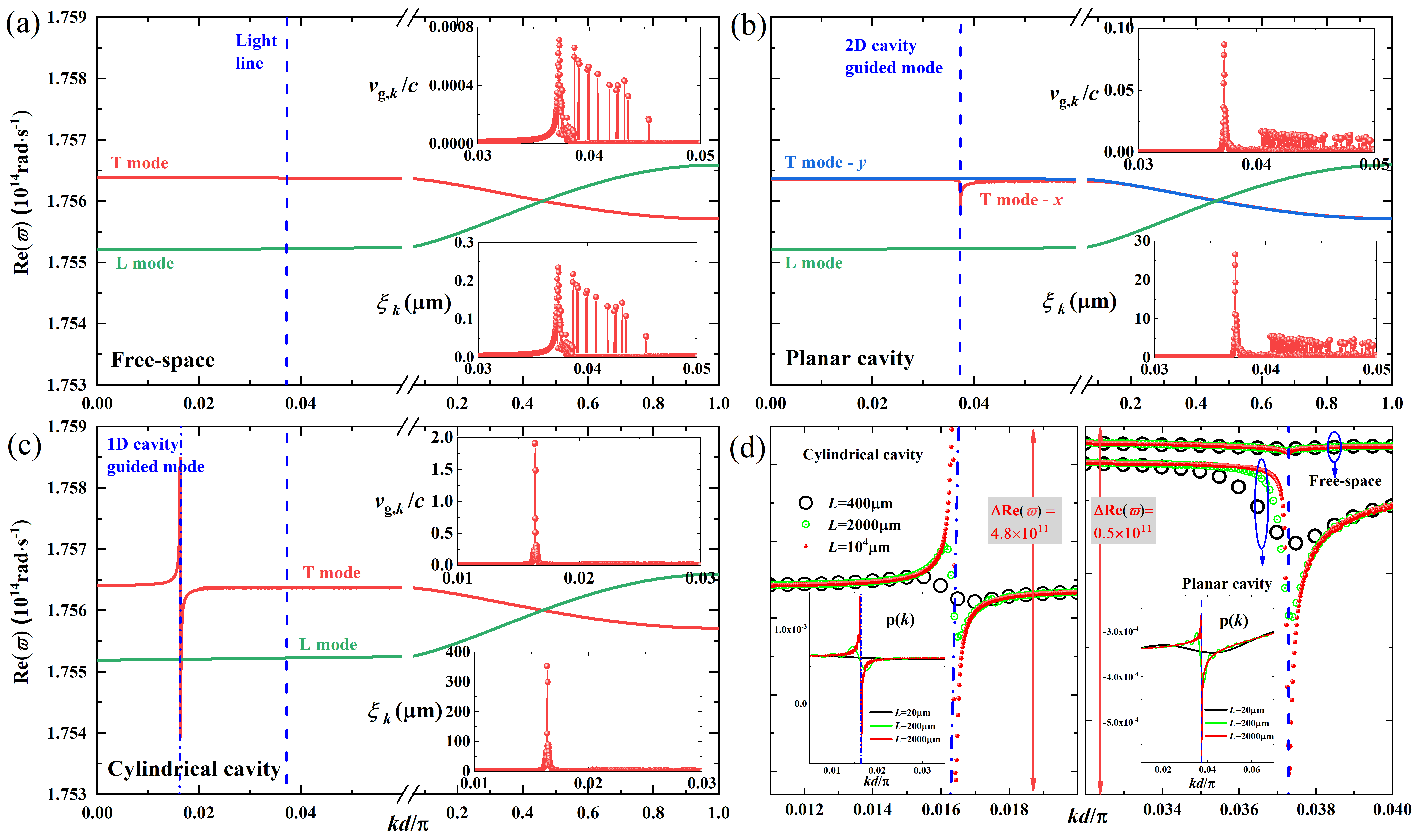}
\caption{\label{}Dispersion relations obtained via eigenfrequency mapping for SiC nanoparticle chains with interparticle spacing $\Lambda = 0.2~\upmu\mathrm{m}$ and total length $L = 10^4~\upmu\mathrm{m}$ ($N = 50{,}001$) in (a) free space, (b) planar cavity and (c) cylindrical cavity. Here, T denotes the transverse modes ($x$- and $y$-polarized excitations) and L the longitudinal mode ($z$-polarized excitation). The insets show the normalized group velocity $v_{\mathrm{g},k}/c$ and the propagation length $\xi_k$ near the light line and within the coupling region. (d) Distribution of the fast eigenmodes for chains of varying lengths.}
\end{figure*}

Figure 2(b) shows the transmission coefficient $\mathcal{T}_{ij}$ between the central NP and NPs at different positions along the chain. At both near- and far-field distances, RHT is dominated by localized surface phonon polaritons (LSPhPs), as indicated by resonant peaks near $\omega=1.756\times10^{14}\,\mathrm{rad/s}$. For $z=0.2\,\upmu\mathrm{m}$, $\mathcal{T}_{ij}$ is similar across all three configurations. At larger separations, however, $\mathcal{T}_{ij}$ in the low-dimensional cavities is enhanced by several orders of magnitude, and in the 1D cavity it remains nearly constant even at long distances. The insets display the spatial distribution of the electric component of the localized density of states $\rho_{\mathrm{E}}$ \cite{Suppl,Joulain2005} when only the central NP thermally emits. LD cavities exhibit an extended range of high $\rho_{\mathrm{E}}$, with the 1D cavity maintaining strong fields along the entire cavity axis.

To gain a deeper understanding of the mechanisms underlying the superballistic regime, we perform a spectral analysis of the system. Since radiative heat transfer (RHT) is dominated by the resonant modes of the SiC NPs, the interparticle coupling in a finite chain can be accurately modeled as a system of coupled harmonic oscillators. Using temporal coupled-mode theory \cite{Haus1984,FanS2003} and denoting $A_i$ as the amplitude of the resonant mode of the $i$th NP, the dynamic of the many-body system is governed by
\begin{equation}
\frac{dA_i}{dt} = -i\omega_0 A_i - (\gamma_{\text{a}} + \gamma_{\text{e}}) A_i - \sum_{j\neq i} i k_{ij} A_j + \sqrt{2\gamma_{\text{e}}}\, s_i^+,
\end{equation}
where $\omega_0$ is the resonant frequency, $\gamma_{\text{a}}$ and $\gamma_{\text{e}}$ denote the decay rates due to absorption and radiation, $k_{ij}$ is the interparticle coupling coefficient, and $s_i^+$ is the incident wave amplitude. These parameters can be extracted by fitting the coupled dipole equations for the NP chain \cite{Suppl}.  
The complex eigenfrequencies $\varpi$ (normal mode frequencies) are obtained by solving the eigenvalue problem
\begin{equation}
i(\mathbf{H} - \varpi)\mathbf{A} = 0,
\end{equation}
where the Hamiltonian matrix $\mathbf{H}$ has elements
\begin{equation}
H_{ij} = \delta_{ij} \big[ \omega_0 - i (\gamma_{\text{a}} + \gamma_{\text{e}}) \big] + (1-\delta_{ij}) k_{ij}.
\end{equation}
For a periodic chain of $N$ oscillators, $N$ discrete eigenfrequencies exist, corresponding to allowed wavevectors $k = n\pi / N$ with $n = 0,1,\dots,N-1$. Each eigenmode of the finite chain is then mapped onto an $\omega\text{-}k$ diagram by assigning a wavevector based on its oscillation pattern \cite{Suppl}.  
Additionally, the dispersion relations of the cavity-guided modes are computed \cite{Suppl}, which can be viewed as freely propagating modes or as the “light line” in low-dimensional spaces. Eigenmodes above the light line correspond to radiative modes, while those below represent evanescent guided modes along the chain.

Figure~3(a)-(c) present the dispersion relations of chains composed of 50{,}001 SiC nanoparticles for the three configurations under consideration.
In free space, the supported modes display relatively flat dispersion curves, leading to low group velocities $v_{\mathrm{g},k}/c$ (see inset of Fig.~3(a)) and, consequently, inefficient energy propagation along the chain.
In contrast, in the planar cavity configuration, the dispersion of the $x$-polarized transverse mode shows a clear coupling with the fundamental cavity mode over a narrow spectral range, while the other modes remain essentially unaffected, similarly to the free-space case. This coupling significantly enhances the group velocity of the transverse mode (inset of Fig.~3(b)), thereby promoting more efficient heat transport along the chain.
Finally, for the cylindrical cavity, the transverse mode exhibits an asymmetric splitting around the guided cavity mode over a much broader frequency range. The resulting anticrossing behavior is a clear signature of strong coupling with the guided mode, which dramatically increases the group velocity (inset of Fig.~3(c)). Notably, in this spectral region, the group velocity even becomes superluminal. However, obviously this does not allow for superluminal energy transport.
On the other hand, the propagation length
\begin{equation}
\xi_k = -\left| v_{\mathrm{g},k} \right| \left[ 2\,\mathrm{Im}\!\left( \varpi \right) \right]^{-1}
\end{equation}
of the transverse mode near these steep dispersion regions is also significantly enhanced. It increases by approximately one order of magnitude in the planar cavity and by two orders of magnitude, reaching hundreds of micrometers, in the cylindrical waveguide. This substantial improvement facilitates long-range energy transport mediated by resonant coupling between the chain modes and the guided cavity mode.
Furthermore, the dispersion relation of the resonant modes, as well as their coupling to the guided modes, exhibits a strong dependence on the system size. Figure~3(d) presents a magnified view of the fast eigenmodes in the coupling region for both the planar and cylindrical cavity configurations, shown for different chain lengths $L$. As $L$ increases, the hybridization between the resonant nanoparticle modes and the cavity-guided modes becomes both stronger and spectrally broader, leading to steeper dispersion branches and a larger number of accelerated eigenmodes. This size-dependent enhancement of the coupling is particularly pronounced in the one-dimensional (cylindrical) cavity configuration.

\begin{figure}
\includegraphics[width=\columnwidth]{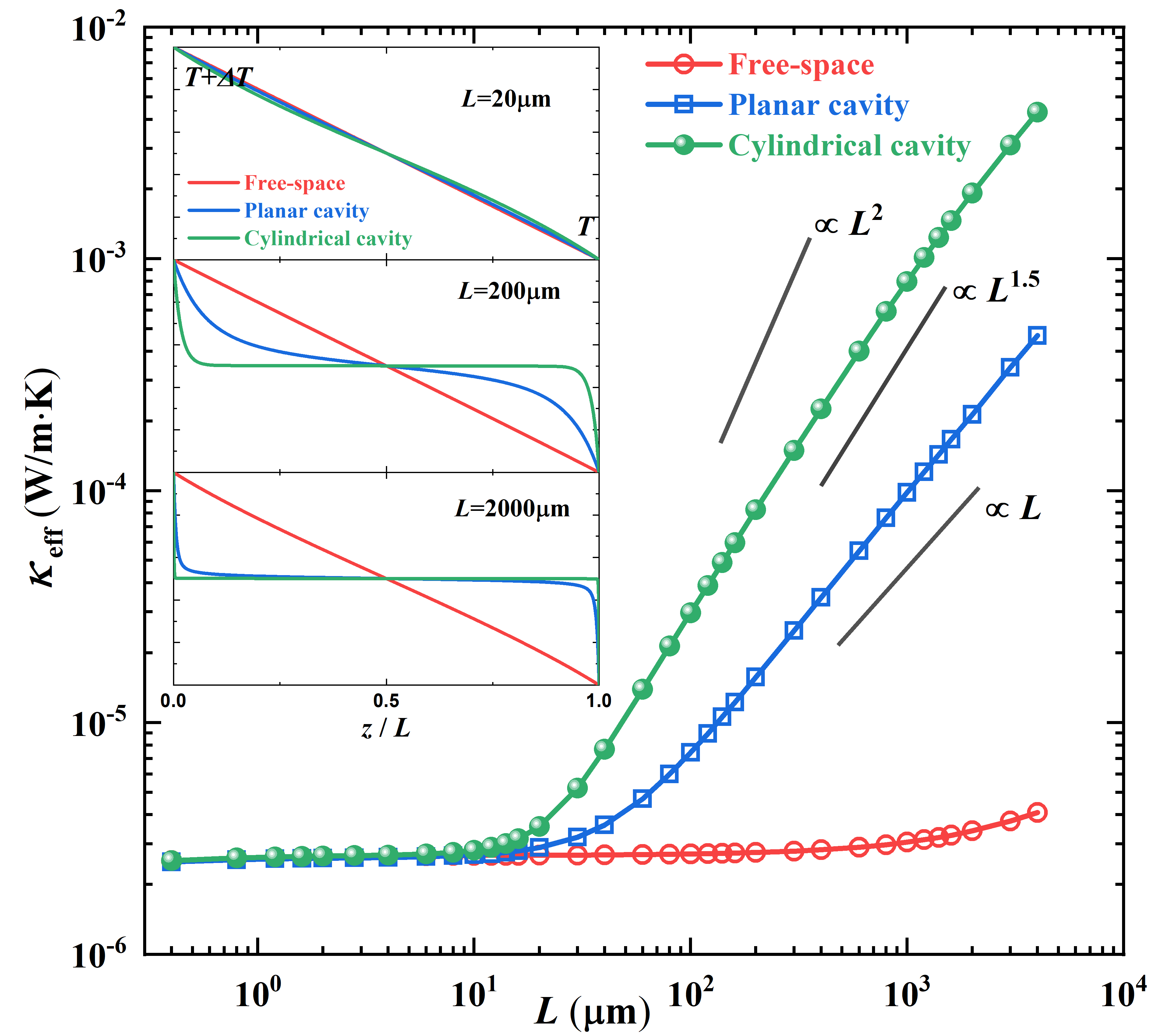}
\caption{Scaling of the effective thermal conductivity $\kappa_{\text{eff}}$ at $T=300\,K$ of free and confined SiC nanoparticle chains ($R=25\,\mathrm{nm}$) with lattice constant $\Lambda=0.2\upmu$m versus the chain length $L$. The  gap size of the planar cavity is $h=1\,\upmu\mathrm{m}$ and the diameter of cylinder is $D=7\,\upmu\mathrm{m}$. Inset:   temperature profiles along the chains in steady state regime.}
\end{figure}
Finally, we examine the size dependence of the effective thermal conductivity $\kappa_{\text{eff}}$, by calculating the heat flux $Q(T)$ crossing the system under an applied temperature gradient $\nabla T$ using the phenomenological Fourier’s law (see detail in \cite{Suppl}) 
\begin{equation}
Q(T)=-\kappa_{\text{eff}}(T) \nabla T.
\end{equation} 
Although classical kinetic theory links heat transport to mode group velocities and propagation lengths of modes, this approach generally fails to capture radiative fluxes in many-body systems \cite{Kathmann2018}. Here, we rigorously calculate the heat flux using the many-body RHT theory \cite{PBA2011}.  
Figure 4 shows $\kappa_{\text{eff}}$ as a function of chain length $L$. For $L \lesssim 10\,\upmu\mathrm{m}$, $\kappa_{\text{eff}}$ is nearly constant for all three cases, dominated by near-field RHT. As $L$ increases, the planar cavity exhibits linear scaling, while the cylindrical cavity shows superlinear growth with an approximate exponent of 1.5, providing clear evidence of ballistic and superballistic transport. In contrast, $\kappa_{\text{eff}}$ in free space grows slowly with $L$. Chains with larger lattice spacing $\Lambda$ support longer ranges of linear and superlinear scaling~\cite{Suppl}.  
The inset of the figure shows steady-state temperature profiles along the chain. For short chains, the profiles are nearly linear, indicating diffusive or quasi-diffusive (short range) transport. For longer chains, however, flat temperature plateaus appear in the interior, accompanied by sharp temperature jumps at the boundaries in the cavity cases. This plateau corresponds to a Casimir-like regime, where the chain interior equilibrates via collective long-range interactions, while energy exchange occurs primarily at the contacts. The 1D cavity exhibits the longest plateau and the steepest boundary jumps, providing additional evidence of anomalous, long-range heat transport.

To conclude, we have shown that radiative heat transport in confined chains of plasmonic nanoparticles can reach the ballistic regime even in sparse chains. In cylindrical cavities, transport can further exceed the classical ballistic limit, entering a superballistic regime. This behavior originates from the cooperative amplification of long-range interactions mediated by cavity-guided modes, which generate accelerated, spatially delocalized transport eigenmodes and produce a superlinear scaling of the effective thermal conductivity with system size. These findings provide a framework for ultrafast photonic heat transport and offer a versatile platform to control energy flow at the nanoscale, with potential applications in thermal management, information processing and quantum technologies.
\begin{acknowledgments}
The supports from National Natural Science Foundation of China (No. 51976045 and 51906128) and project ZR2025MS768 supported by Shandong Provincial Natural Science Foundation are acknowledged.
\end{acknowledgments}
\end{document}